\documentclass
[twocolumn,showpacs,preprintnumbers,amsmath,amssymb,prd,nofootinbib,superscriptaddress]{revtex4}%
\usepackage{bm}
\usepackage{hyperref}
\usepackage{mathrsfs}
\usepackage{xcolor,color,graphicx,graphics}
\usepackage[all]{xy}
\usepackage{epsfig,subfigure}
\usepackage{latexsym,amssymb,amsmath,amsfonts} 
\usepackage[english]{babel} 
\usepackage[OT1]{fontenc}
\usepackage[latin1]{inputenc}
\usepackage{makeidx}
\usepackage{hyperref}
\usepackage{color,graphicx,graphics,wrapfig,epsf}
\usepackage{caption}
\usepackage{mwe}

\hypersetup{urlcolor=BlueViolet,
	    citecolor=Plum,
	    linkcolor=PineGreen}

\begin{document}

\title{Metric-affine gravity effects on terrestrial (exo-)planets profiles}

\author{Aleksander Kozak}
\email{aleksander.kozak@uwr.edu.pl}
\affiliation{Institute of Theoretical physics, University of Wroclaw, pl. Maxa Borna 9, 50-206 Wroclaw, Poland
}

\author{Aneta Wojnar}
\email{aneta.magdalena.wojnar@ut.ee}
\affiliation{Laboratory of Theoretical Physics, Institute of Physics, University of Tartu,
W. Ostwaldi 1, 50411 Tartu, Estonia
}

\begin{abstract}
Mass-radius relations of homogeneous cold spheres are obtained for six solid materials commonly found in terrestrial planets. An additional degeneracy in the (exo-)planets' profiles is discussed together with their properties concluded from our findings in the framework of Palatini $f(\mathcal R)$ gravity. Moreover, a new test of gravity has been proposed: The results presented here will allow to test and to constrain models of gravity by the use of seismic data acquired from earthquakes and marsquakes.

\end{abstract}

\maketitle

\section{Introduction}

General Relativity (GR) has been tested experimentally on a number of occasions \cite{Will:2014kxa}. Just recently, the existence of one of the most astonishing predictions made by Einstein's theory, black holes, has been confirmed by detecting gravitational waves coming from a merger of two of them \cite{TheLIGOScientific:2017qsa}. Also, the shadow of a supermassive black hole in the center of M87 galaxy has been recently observed directly \cite{Akiyama:2019cqa,aki2,aki3,god}
(see \cite{Barack:2018yly} for a review). Despite all these triumphs, GR  cannot account for various cosmological and astrophysical phenomena in a satisfactory way. A lot of effort has been dedicated to constructing alternative models being capable of solving dark matter and dark energy problem \cite{Copeland:2006wr,Nojiri:2006ri,nojiri2,nojiri3,Capozziello:2007ec,Carroll:2004de}, sheding some light on space-time singularities \cite{Senovilla:2014gza}, or providing unification scenarios at high energies \cite{ParTom,BirDav}. Another problem concerns the observed maximum masses of compact objects that exceed theoretical predictions \cite{lina,as,craw,NSBH}, and mass of a binary black hole merger \cite{abotHBH,sak3}.

In this work we use the fact that many models of gravitation, in particular $f(\mathcal{R})$ Palatini gravity (see the section \ref{palatini} for a short review of the model), slightly alter the non-relativistic limit of (sub-)stellar structural equations by introducing new (geometric) terms proportional to functions of energy density \cite{Saito:2015fza,olek,olmo_ricci} (for a review see \cite{review,cantata}). Modified non-relativistic equations in the context of stars and brown dwarfs have been already widely used by the physics community, mainly to obtain limiting masses, such as e.g., the Chandrasekhar mass for white dwarf stars \cite{Chandra,Saltas:2018mxc,Jain:2015edg,Banerjee:2017uwz,Wojnar:2020wd,Belfaqih:2021jvu}, the minimum Main Sequence mass\footnote{that is, a star reaches the Main Sequence when the energy produced in the star's core by hydrogen burning is balancing the gravitational contraction.} \cite{Sakstein:2015zoa,Sakstein:2015aac,Crisostomi:2019yfo,Olmo:2019qsj}, or minimum mass for deuterium burning  \cite{Rosyadi:2019hdb}. Moreover, modified gravity also impacts the early evolution of low-mass stars \cite{early}, and the post-Main Sequence stage of population II stars \cite{chow}, cooling processes of brown dwarfs \cite{Benito2021}, as well as it alters the age-estimation procedures based on the lithium depletion method \cite{Wojnar:2020frr}.

Therefore, using modified equations describing a spherical-symmetric object, we reveal an additional degeneracy induced by metric-affine gravity in the mass-radius relations for a cold homogeneous sphere. Such an object can be treated as a single-layer (exo-)planet and it is useful to demonstrate that the modified gravity effects take also part in the planetary description. To show that, we use equations of state in analytical form for six solid materials, and modified hydrostatic equilibrium equations presented in the section \ref{secplanets} and appendix \ref{appen}. We also discuss a possible singularity caused by a particular combination of an equation of state and the theory parameter, and we argue that for a physical system such as a star or a planet, this problem will not appear. In the section \ref{num} we numerically solve the equations and demonstrate the mass-radius relations and density profiles obtained for the Palatini quadratic model. The section \ref{test} is devoted to a description of a new test of gravity with the use of our results and seismic data from earthquakes and marsquakes. In the last section we draw our conclusions. 

Let us notice that so far, the (exo-)planets have been used to test and constrain theories of gravity only in the context of precessions of planetary perihelion in the Solar System, see e.g. \cite{BD,melinda,iorio,harko,hatice,bag,seba,schmidt,harko2,iorio2,fatibene,bhat}, and modifications to the third Kepler's law by introducing corrections to the third Kepler's law \cite{kepler1,kepler2}.

We use $(-+++)$ signature convention and $\kappa^2=8\pi G/c^4$.

\section{Palatini $f(\mathcal{R})$ gravity}\label{palatini}

In the metric approach to $f(R)$ gravity, where one replaces Einstein-Hilbert Lagrangian with a general function of the curvature scalar, metric tensor is the only object mediating the gravitational interaction. It is possible, however, to introduce an independent connection and split up spacetime structure into metric and affine. Such an approach is called 'Palatini', and it exhibits some advantages over the metric formulations \cite{junior,sch,ol1,ol2}.

The action is given by:

\begin{equation}
S[g,\Gamma,\psi_m]=\frac{1}{2\kappa^2}\int \sqrt{-g}f(\mathcal{R}) d^4 x+S_{\text{matter}}[g,\psi_m],\label{action}
\end{equation}
where $\mathcal{R} = g^{\mu\nu}\mathcal{R}_{\mu\nu}(\Gamma)$ is the Palatini curvature scalar, built both from the metric and the independent connection, and $\psi_m$ represents matter fields.

Varying the action with respect to the metric tensor yields:

\begin{equation}
f'(\mathcal{R})\mathcal{R}_{\mu\nu}-\frac{1}{2}f(\mathcal{R})g_{\mu\nu}=\kappa^2 T_{\mu\nu},\label{structural}
\end{equation}
with $T_{\mu\nu}=-\frac{2}{\sqrt{-g}}\frac{\delta S_m}{\delta g_{\mu\nu}}$ and prime denoting differentiating with respect to the curvature. One can contract the equation \eqref{structural} with the metric and relate the Palatini curvature to the trace of energy-momentum tensor:

\begin{equation}
    f'(\mathcal{R})\mathcal{R}-2f(\mathcal{R})=\kappa^2 T.
\end{equation}
We immediately notice that the relation between the curvature and the trace becomes purely algebraic, which means that, for a particular choice of the function $f$, it is possible to solve it. 

In order to obtain relation between the connection and the metric tensor, one needs to vary with respect to it and obtain the following:
\begin{equation}
\nabla_\beta(\sqrt{-g}f'(\mathcal{R}(T))g^{\mu\nu})=0,\label{con}
\end{equation}
where the covariant derivative is defined using the independent connection. If one defines a new metric tensor, conformally related to $g_{\mu\nu}$:
\begin{equation}
    h_{\mu\nu} = f'(\mathcal{R}(T))g_{\mu\nu}
\end{equation}
then the equation \eqref{con} can be written as:
\begin{equation}
\nabla_\beta(\sqrt{-h}h^{\mu\nu})=0,\label{con2}
\end{equation}
and by a well-known theorem, it means that the connection $\Gamma^\alpha_{\mu\nu}$ is Levi-Civita with respect to the metric $h_{\mu\nu}$  \cite{DeFelice:2010aj,BSS,SSB}. Therefore, the 'independent' connection turns out to be an auxiliary field and can be integrated out. All relevant degrees of freedom are related to the metric tensor.

Just like its metric counterpart, Palatini $f(\mathcal{R})$ has a scalar-tensor representation, which can be obtained by means of the Legendre transformation:

\begin{equation}
\begin{split}
S[g,\Phi,\Gamma,\psi_m]=&\frac{1}{2\kappa^2}\int \sqrt{-g}[\Phi\mathcal{R} - V(\Phi)]d^4 x\\
&+S_{\text{matter}}[g,\psi_m],    
\end{split}
\label{STPalatini}
\end{equation}
where $\Phi = df/dR$ and $V(\Phi) = f'(\mathcal{R}(\Phi))\mathcal{R}(\Phi) - f(\mathcal{R}(\Phi))$. It can significantly simplify the problems analyzed \cite{aneta_stab,aneta,o,o1,o2}. The connection, building the Ricci tensor, can be expressed in terms of the scalar field $\Phi$ and the metric tensor $g_{\mu\nu}$, yielding:
\begin{equation}
\begin{split}
S[g,\Phi,\psi_m]=&\frac{1}{2\kappa^2}\int \sqrt{-g}\left[\Phi R +\frac{3}{2\Phi}(\partial\Phi)^2- V(\Phi)\right]d^4 x \\
&+S_{\text{matter}}[g,\psi_m],    
\end{split}
\label{STPalatini2}
\end{equation}
which is, effectively, a fully-metric theory. 

Palatini $f(\mathcal{R})$ theory is a special case of a more general class of modified gravity, represented by the following action functional:

\begin{equation}
    \begin{split}
        S[g,\Phi,\Gamma,\psi_m]=&\frac{1}{2\kappa^2}\int \sqrt{-g}\Big[\mathcal{A}(\Phi) R(g, \Gamma) -\mathcal{B}(\Phi)(\partial\Phi)^2\\
       & - \mathcal{V}(\Phi)\Big]d^4 x
+S_{\text{matter}}[e^{2\alpha(\Phi)}g,\psi_m].
    \end{split}
\end{equation}

Here, in order to keep the considerations at the most general level possible, we do not specify if the curvature is fully metric, or constructed \'{a} la Palatini. As one can see, there are four functions of the scalar field, entering the action. Specifying the functions, one gets a particular scalar-tensor theory. By means of field equations, one can establish an equivalence between Palatini and metric approaches to this class of modified gravity; such representations will have different values of the scalar field functions. 

For any scalar-tensor theory, either in the metric, or in the Palatini formalism, one can introduce quantities whose value remains the same under a Weyl (or conformal) transformation of the metric tensor and a re-parametrization of the scalar field, defined by \cite{borow2020, akab, kuusk22015, tartu, kuusk2016, karam2017, karam2018}:

\begin{eqnarray}
\begin{cases}
\bar{g}_{\mu\nu} = e^{2\gamma(\Phi)}g_{\mu\nu} \\
\bar{\Phi} = \bar{f}(\Phi)
\end{cases}
\end{eqnarray}
where $\gamma(\Phi)$ is some function of the scalar field. Performing a conformal change might be treated as a mathematical tool allowing one to choose a set of scalar field functions, for which solving field equations will be particularly simple; expressing relevant quantities in terms of invariants might provide framework for analysis of different approaches and theories within one framework. 

The invariants used in this work are defined as follows:
\begin{eqnarray}
\mathcal{I}_1(\Phi)& =& \frac{\mathcal{A}(\Phi)}{e^{2\alpha(\Phi)}}, \\
\mathcal{I}_2(\Phi)& =& \frac{\mathcal{V}(\Phi)}{\mathcal{A}^2(\Phi)},\\
\frac{d\mathcal{I}}{d\Phi}& =& \sqrt{\frac{\mathcal{B}}{\mathcal{A}}+\delta_\Gamma \left(\frac{3\mathcal{A}'}{2\mathcal{A}}\right)^2},
\end{eqnarray}
where $\delta_\Gamma$ is one for metric theory, and zero for Palatini. Prime denotes differentiating with respect to the scalar field.

\section{Palatini planets}\label{secplanets}

\subsection{Equations of state for cold low-mass spheres}
In the following work we will use equations of state for six different solid materials (see table \ref{tabbirch} and \ref{tabpoly}). The first one concerns a case when one deals with the assumption on uniform or zero temperature, with pressures below $200\,\text{GPa}$. For such conditions, we are equipped with the analytical form of the EoS given by the fits to the experimental data:
\begin{equation}\label{BME}
    p=\frac{3}{2}K_0(\eta^{7/3}-\eta^{5/3})\left( 1+\frac{3}{4}(K'_0-4)(\eta^{2/3}-1) \right),
\end{equation}
where $\eta=\rho/\rho_0$ is the compression ratio with respect to the ambient density $\rho_0$ (that is, the density at zero pressure), $K_0=-V(\partial p/\partial V)_T$ is the bulk modulus of the material (the inverse of the compressibility) \cite{wep}, while $K'_0$ and $K''_0$ are the first and second pressure derivatives, respectively. Since most of the experiments are limited to $p<150\,\text{GPa}$ and temperature less than $2000$K, we will take it as the starting value at the core of the planet.
The above EoS is called the third-order finite strain Birch-Murgnagham equation of state (BME) \cite{birch, poi}. In the table \ref{tabbirch} there are only two materials which we are using in this work - see our discussion in the section \ref{num}; for more fits for various materials, see e.g. \cite{poi} and \cite{seager}.

\begin{table}
\caption{Best-fit parameters for BME (\ref{BME}); for more materials, see e.g. the table 4.1 in the reference \cite{poi} and the table 1 in the reference \cite{seager}.}
\centering
\begin{tabular}{llll }
\hline\noalign{\smallskip}
Material & $\rho_0$ (\text{Mg m}$^{-3}$) & $K_0$ (\text{GPa})  & $K_0'$ \\
\noalign{\smallskip}\hline\noalign{\smallskip}
Fe($\alpha$) & 7.86 & 162.5 & 5.5\\
(Mg, Fe)SiO$_3$ & 4.26 & 266 & 3.9 \\
\noalign{\smallskip}\hline
\end{tabular}\label{tabbirch}
\end{table}

For $p\gtrsim10^4\,\text{GPa}$, the electron degeneracy becomes important. The common approach is to match the equation (\ref{BME}) with the Thomas-Fermi-Dirac EoS (TFD) \cite{thomas, fermi, dirac, feynmann} with a density-dependent correlation energy term added \cite{sal} in order to take into account interactions between electrons themselves, obeying the Pauli exclusion principle and  moving in the Coulomb field of the nuclei. 
However, it turns one that the merger of the BME and TFD equations of state can be approximated by a modified polytropic equation of state \cite{seager}, called further Seager-Kuchner-Hier-Majumder-Militzer EoS (SKHM EoS),
\begin{equation}\label{pol}
    \rho(p)=\rho_0 +cp^n,
\end{equation}
whose best-fit parameters $\rho_0$, $c$, and $n$ are given in the table \ref{tabpoly}. The reason of such a modification, given here by the added $\rho_0$, is to include the incompressibility of solids and liquids at low pressures. The equation (\ref{pol}) with the given fits for the considered solid materials is valid for the pressure range $p<10^{7}\,\text{GPa}$.

\begin{table}
\caption{Best-fit parameters for the SKHM EoS (\ref{pol}) obtained in the reference \cite{seager}.}
\centering
\begin{tabular}{llll }
\hline\noalign{\smallskip}
Material & $\rho_0$ (\text{kg m}$^{-3}$) & $c$ (\text{kg m}$^{-3}$ \text{Pa}$^{-n})$ & $n$ \\
\noalign{\smallskip}\hline\noalign{\smallskip}
Fe($\alpha$) & 8300 & 0.00349 & 0.528\\

 MgSiO$_3$ & 4100 & 0.00161 & 0.541 \\

 (Mg, Fe)SiO$_3$ & 4260 & 0.00127 & 0.549 \\
 
 H$_2$O (ice) & 1460 & 0.00311 & 0.513 \\
  
 C (graphite)  & 2250 & 0.00350 & 0.514 \\
   
 SiC & 3220 & 0.00172 & 0.537 \\
\noalign{\smallskip}\hline
\end{tabular}\label{tabpoly}
\end{table}

\subsection{General structural equations}\label{alfa}

It was shown that the full relativistic hydrostatic equilibrium equation for quadratic model in Palatini $f(\mathcal R)$ gravity:
$$
f(\mathcal R)=\mathcal R + \beta \mathcal R^2
$$
is given by \cite{aneta_stab, olek}:
\begin{align}\label{tovJ}
        p' =  \Bigg[&-\frac{G\mathcal{M}(r)}{c^2r^2\mathcal{I}^{1/2}_1}(c^2\rho+p)\left(1 - \frac{2G\mathcal{M}(r)}{c^2r\mathcal{I}^{1/2}_1}\right)^{-1}\nonumber\\
        &\times\left(1 + \frac{4\pi\mathcal{I}_1^\frac{3}{2}r^3}{c^2\mathcal{M}(r)}\left(\frac{p}{\mathcal{I}^2_1} + \frac{\mathcal{I}_2}{2\kappa^2}\right)\right)\Bigg]  \\
        &\times \left(\frac{r}{2}\partial_{r} \ln\mathcal{I}_1 + 1 \right)
   +\left(-c^2\rho + 5p\right)\partial_{r} \ln\mathcal{I}_1,\nonumber
\end{align}
where the invariants in the case of perfect-fluid energy-momentum tensor are:
\begin{align}
    & \mathcal{I}_1 =  1 + 4\beta \kappa^2 (c^2\rho - 3p), \\
    & \mathcal{I}_2 = \frac{4\beta\kappa^4(c^2\rho - 3p)^2}{\left(1 + 4\beta \kappa^2 (c^2\rho - 3p) \right)^2},
\end{align}
and prime denotes the derivative with respect to the $r$ coordinate.
The mass function, after applying the above forms of the invariants, is:
\begin{equation}\label{masaRel}
\begin{split}
    \mathcal{M}(r) = &   \int^{r}_0 4\pi\tilde{r}^2 \frac{\rho - 2c^{-2}\beta\kappa^2(c^2\rho - 3p)^2}{\left(1 + 4\beta \kappa^2 (c^2\rho - 3p) \right)^{1/2}}\\
    &\times\left[1+ \frac{\tilde{r}}{2}\partial_{\tilde{r}}\ln\left(1+4\beta\kappa^2 (c^2\rho -3p)\right)\right]d\tilde{r}.
\end{split}
\end{equation}
Let us notice that the above equations are exact; that is, no approximation has been applied yet. 
In what follows, we will use a redefined Starobinsky parameter as:
\begin{equation}
    \alpha := 2 c^2 \kappa^2 \beta.
\end{equation}
It should be commented that such defined parameter has a very small value in contrary to the Starobinsky parameter $\beta$, whose value must be large in order to take into account the higher curvature term, given here by $\mathcal R^2$.

Since the conformal transformation (given by the invariant $\mathcal{I}_1$ in our model) was used in order to get the equations written with respect to the physical quantities, the mass and hydrostatic equilibrium equations are singular for a particular value of the parameter $\alpha$ which depends on energy density and pressure:
\begin{equation}\label{singu}
    \alpha_\text{sing}=-\frac{1}{2(\rho-\frac{3p}{c^2})}.
\end{equation}

A possible singular behaviour of the additional terms in TOV equations provided by particular equations of state in different models of gravity has been already detected in previous works \cite{ray, olmo,kim, buch1, buch2,astab}. Such a feature arises from the fact that the modifications often provide new matter-dependent contributions to the hydrostatic equilibrium equation and stability conditions \cite{aneta_stab, early}, via algebraic relation between scalar curvature and matter, as it happens in models of metric-affine gravity considered here, or modified Klein-Gordon equation, relating the dynamics of the scalar field\footnote{which is an extra degree of freedom in a given scalar-tensor theory} with the ordinary matter sources.

Moreover, it should be taken into account that the singular value of the parameter changes with the energy density and pressure, according to their profiles; that is, it will have much smaller value in the core than nearby the object's surface. Therefore, one should be always very careful when choosing a particular negative value of the parameter $\alpha$, taking into account given equation of state and the range of densities/pressures through an examined object. In our case we are not even close to the singular value as we would have to consider densities much higher than the ones present in the planetary interiors (and together with the considered EoS's describing planetary compositions) in order to have parameter's values making the equations singular.

\subsection{Structural equation for terrestrial planets}

Since we are dealing with planets, the terms proportional to $p/c^2$ are negligible when compared to the energy density, and hence the equation (\ref{tovJ}) reduces to:
\begin{equation}\label{tovJ1}
    \begin{split}
        & p' =  -\frac{G\mathcal{M} \rho}{r^2\mathcal{I}^{1/2}_1}\left(1 - \frac{2G\mathcal{M}}{c^2r\mathcal{I}^{1/2}_1}\right)^{-1}\\
        &\times\left(1 + \frac{4\pi\mathcal{I}_1^\frac{3}{2}r^3}{c^2\mathcal{M}}\frac{\mathcal{I}_2}{2\kappa^2}\right)\left(\frac{r}{2}\partial_{r} \ln\mathcal{I}_1 + 1 \right) \\
        & 
   -c^2\rho \:\partial_{r} \ln\mathcal{I}_1
    \end{split}
\end{equation}
with:
\begin{align*}
    & \mathcal{I}_1 = 1 + 2\alpha\rho ,\;\;\;\;\mathcal{I}_2 = \frac{2\alpha\kappa^2c^2\rho^2}{\left(1 + 2\alpha\rho  \right)^2}. 
\end{align*}
The non-relativistic mass is then:
\begin{equation}\label{mass1}
    \mathcal{M}(r) =    \int^{r}_0 4\pi\tilde{r}^2 \frac{\rho - \alpha\rho^2}{\left(1 + 2\alpha\rho  \right)^{1/2}}
   \left[1+ \frac{\tilde{r}}{2}\partial_{\tilde{r}}\ln\left(1+2\alpha\rho \right)\right]d\tilde{r},
\end{equation}
while the singular value of the parameter $\alpha$ can be approximated as:
\begin{equation}
    \alpha_\text{sing}=-\frac{1}{2\rho}.
\end{equation}
It is not the common non-relativistic limit (we have just neglected the pressure $p$ contribution): the geometric term as well as the gravitational pressure contribution coming from the invariants $\mathcal{I}_1$ and $\mathcal{I}_2$ are still present. The reason for that is when one expands the equations around $\alpha=0$, which as we discussed, will be a very small value in our case, one loses the discussed information about the singular behaviour of the equations related to a particular value(s) of the parameter, and moreover, as we have checked (see the discussion and  equations in the appendix \ref{appen}), the non-physical profiles occur for the positive $\alpha$ in the case of large planets. Because of that fact, we will use the above equations in order to include planets with larger masses, that is, to allow the planets to reach the mass limit when the electron degeneracy must be taken into account.

The structural equations used in the numerical analysis described in the section \ref{num} are given in the appendix \ref{appen}.

\section{Planet profiles and numerics}\label{num}

In order to obtain mass-radius relationship for planets, we integrate the equation \eqref{tovJ1} and the mass relation \eqref{masaRel} (expressed as derivative). The additional information, allowing one to relate energy density and pressure, is provided by a suitable equation of state. The exact form of the equations used for the case of polytropic equation of state can be found in the Appendix \ref{appen}. 

For a given value of the parameter $\alpha$, the equations were integrated using the 4th order Runge-Kutta method. The initial conditions at $R=0$ were $\mathcal{M}(0) = 0$ and $\rho(0) = \rho_c > \rho_0$, and the integration has been carried out for a wide range of possible central densities, starting from little above $\rho_0$ for a given material, and finishing at the point when electron degeneracy must be taken into account. The process of integrating for a particular value of $\rho_c$ ends when the density drops to its ambient value, i.e. $\rho_0$. At this moment, the final mass and radius are read off and a single point is placed in the mass-radius diagram. After obtaining all relevant data, we change the value of $\alpha$ and repeat the procedure.

\subsection{The Birch-Murgnagham equation of state}

Using the BME (\ref{BME}) applied to the hydrostatic equilibrium and mass equations (\ref{tovJ}) and (\ref{masaRel}), respectively, one obtains the radius-mass curves and density profiles with respect to each material. Since this EoS is suitable only for the low pressure regime, the maximal masses should not cross the Earth one, and even the Earth's core ($p=364$ GPa according to the PREM model \cite{prem}) should not be modeled by it. Therefore, the curves just before reaching the Earth's mass are extrapolated and they do not describe a realistic mass-radius relation for the Earth-size and larger planets.

\begin{figure*}[h]
\centering
    \includegraphics[width=0.5\linewidth]{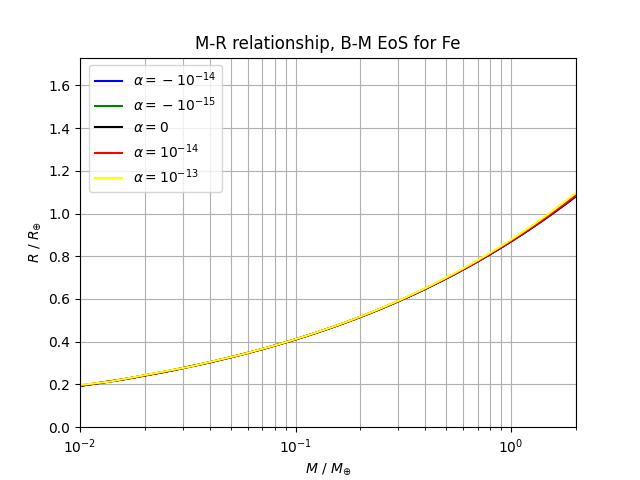}\hfil
    \includegraphics[width=0.5\linewidth]{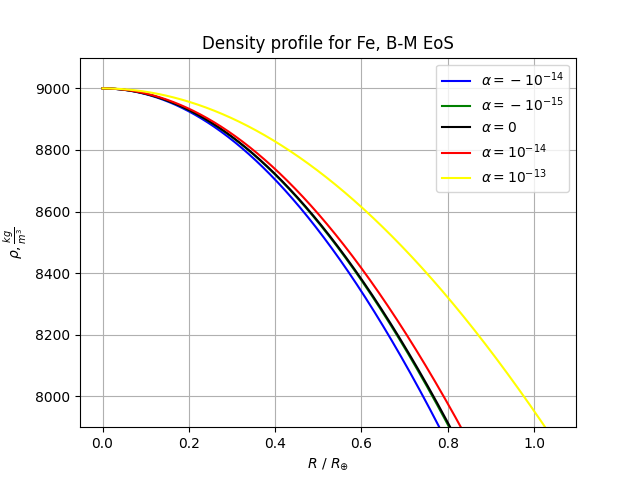}
\caption{[color online] Mass-radius relation and density profile plots for homogeneous cold spheres of iron for different values of the parameter $\alpha$ ($\alpha=0$ gives a GR/Newtonian planet), obtained from the Birch-Murgnagham equation of state with the used of the modified hydrostatic equilibrium equations. }
\label{bm_fe}
\end{figure*}

\begin{figure*}[h]
\centering
    \includegraphics[width=0.5\linewidth]{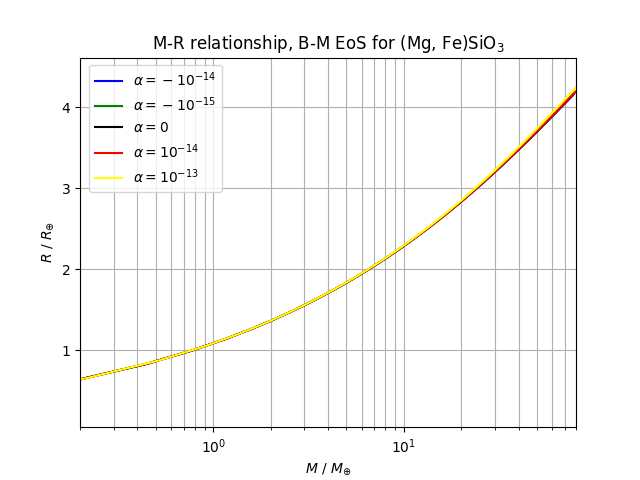}\hfil
    \includegraphics[width=0.5\linewidth]{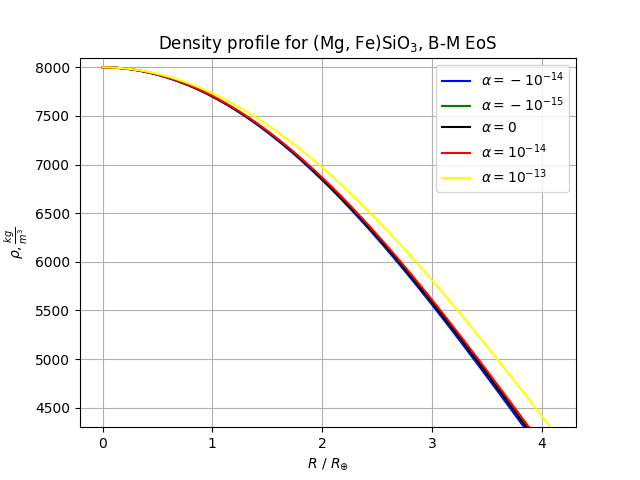}
\caption{[color online]Mass-radius relation and density profile plots for homogeneous cold spheres of silicate (Mg,Fe)SiO$_3$ for different values of the parameter $\alpha$ ($\alpha=0$ gives a GR/Newtonian planet), obtained from the Birch-Murgnagham equation of state with the used of the modified hydrostatic equilibrium equations.}
\label{bm_mgfe}
\end{figure*}

Our analysis demonstrated in the mass-radius plots that the Newtonian and modified gravity curves of the homogeneous cold spheres overlap. Because of that fact, we plotted only the iron and silicates curves (see the figures \ref{bm_fe} and \ref{bm_mgfe}) which are the main abundant materials of the rocky planets in our Solar System. But, let us notice that the density profiles do differ even for so small masses ($\alpha=0$, that is, GR/Newtonian case, is given by the black curve), making possible to use this feature to constrain theories of gravity. We discuss this finding in detail in the section \ref{test}. 

However, as it will be seen in the next subsection, for terrestrial planets with masses and radii bigger than the Solar System ones we will deal with an extra degeneracy in the radius-mass relations provided by modified gravity.

\subsection{The SKHM (modified polytropic) equation of state}

Since the BME (\ref{BME}) does not reproduce reliable results even for the most inner layers of the Earth, we will focus now on the modified polytropic EoS (\ref{pol}), which takes into account the electron degeneracy and interactions between electrons, as well as the particles motion in the Coulomb field of the nuclei. Such an EoS can be used for the pressure range $p<10^{7}\,\text{GPa}$, therefore it is our maximal central value in the numerical approach, expressed by $\rho_c$ obtained from (\ref{pol}).

\begin{figure*}[h]
\centering
    \includegraphics[width=0.5\linewidth]{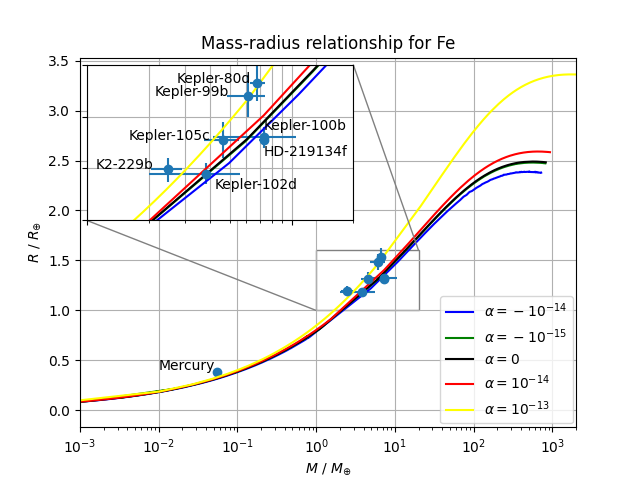}\hfil
    \includegraphics[width=0.5\linewidth]{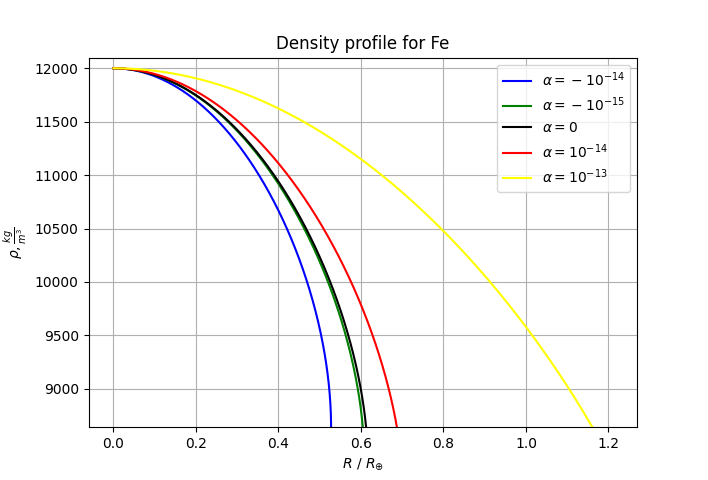}
\caption{[color online] Mass-radius relation and density profile plots for homogeneous cold spheres of iron for different values of the parameter $\alpha$ ($\alpha=0$ gives a GR/Newtonian planet), obtained from the SKHM equation of state with the used of the modified hydrostatic equilibrium equations. Mercury and a few super-Mercury exoplanets are depicted \cite{brugger1, nasa2}.}
\label{fe}
\end{figure*}

\begin{figure*}[h]
\centering
    \includegraphics[width=0.5\linewidth]{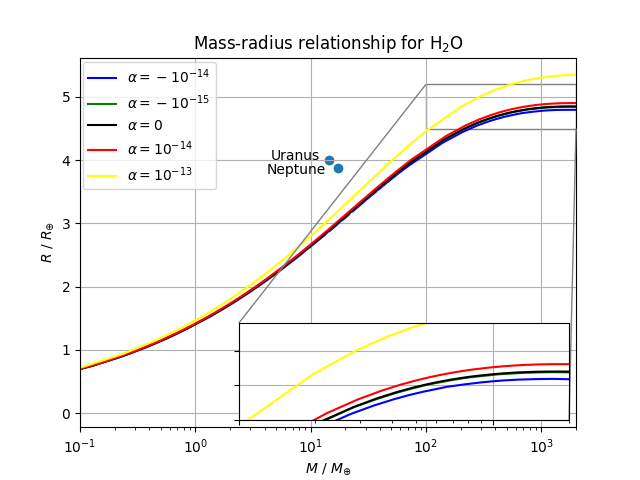}\hfil
    \includegraphics[width=0.5\linewidth]{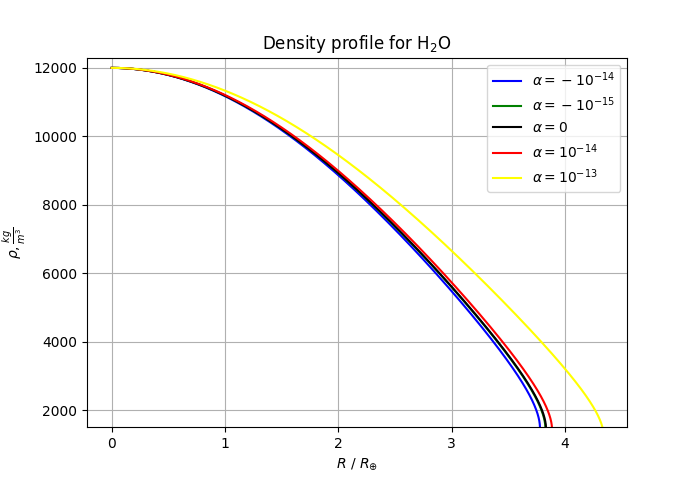}
\caption{[color online] Mass-radius relation and density profile plots for homogeneous cold spheres of water ice for different values of the parameter $\alpha$ ($\alpha=0$ gives a GR/Newtonian planet), obtained from the SKHM equation of state with the used of the modified hydrostatic equilibrium equations. Uranus and Neptune, the giant ice-planets, are depicted.}
\label{h2o}
\end{figure*}

\begin{figure*}[h]
\centering
    \includegraphics[width=0.5\linewidth]{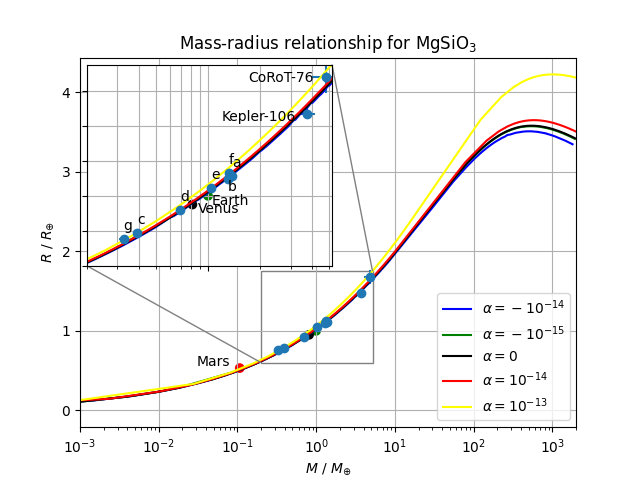}\hfil
    \includegraphics[width=0.5\linewidth]{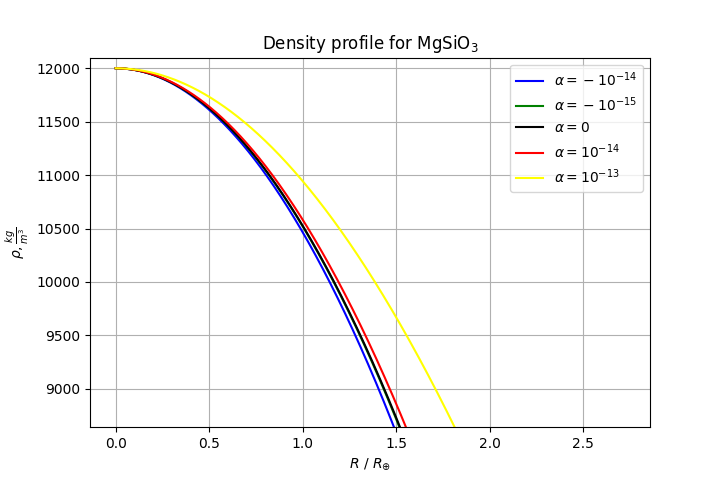}
\caption{[color online] Mass-radius relation and density profile plots for homogeneous cold spheres of silicate MgSiO$_3$ for different values of the parameter $\alpha$ ($\alpha=0$ gives a GR/Newtonian planet), obtained from the SKHM equation of state with the used of the modified hydrostatic equilibrium equations. Mars, Venus and the Earth, as well as a few terrestrial exoplanets are depicted \cite{nasa2, brugger2, ago}. The letters denote TRAPPIST-1  planets, whose physical parameters can be found in \cite{ago}.}
\label{mgsio}
\end{figure*}

\begin{figure*}[h]
\centering
    \includegraphics[width=0.5\linewidth]{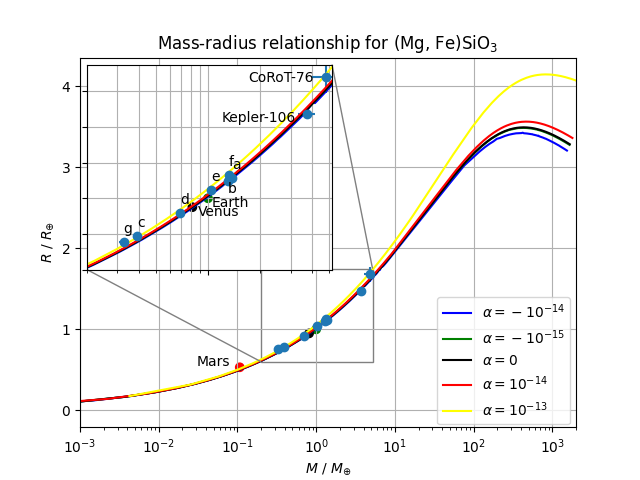}\hfil
    \includegraphics[width=0.5\linewidth]{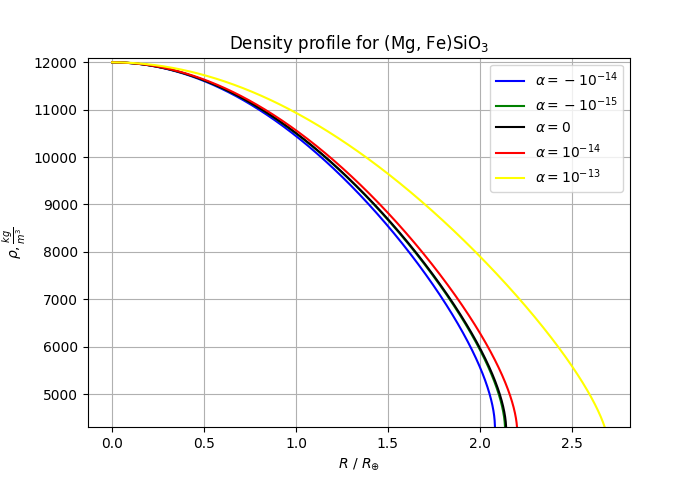}
\caption{[color online] Mass-radius relation and density profile plots for homogeneous cold spheres of silicate (Mg,Fe)SiO$_3$ for different values of the parameter $\alpha$ ($\alpha=0$ gives a GR/Newtonian planet), obtained from the SKHM equation of state with the used of the modified hydrostatic equilibrium equations. Mars, Venus and the Earth, as well as a few terrestrial exoplanets are depicted \cite{nasa2, brugger2, ago}. The letters denote TRAPPIST-1  planets, whose physical parameters can be found in \cite{ago}.}
\label{mgfe}
\end{figure*}

\begin{figure*}[h]
\centering
    \includegraphics[width=0.5\linewidth]{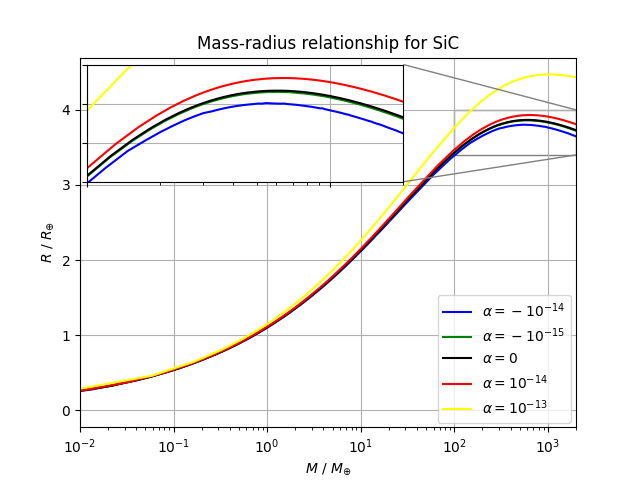}\hfil
    \includegraphics[width=0.5\linewidth]{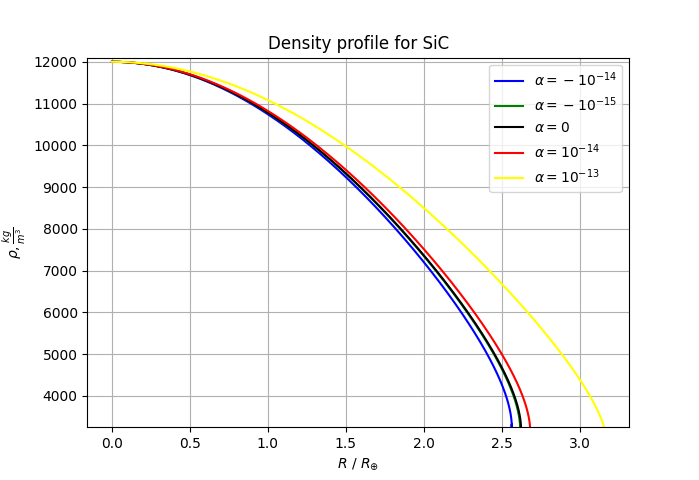}
\caption{[color online] Mass-radius relation and density profile plots for homogeneous cold spheres of silicon carbide for different values of the parameter $\alpha$ ($\alpha=0$ gives a GR/Newtonian planet), obtained from the SKHM equation of state with the used of the modified hydrostatic equilibrium equations. }
\label{sic}
\end{figure*}

\begin{figure*}[h]
\centering
    \includegraphics[width=0.5\linewidth]{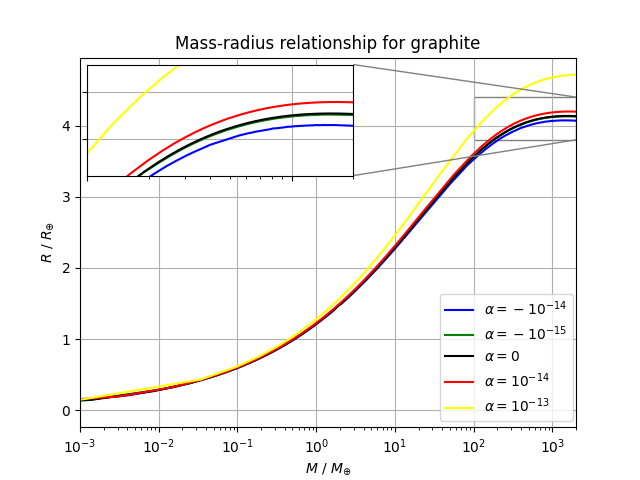}\hfil
    \includegraphics[width=0.5\linewidth]{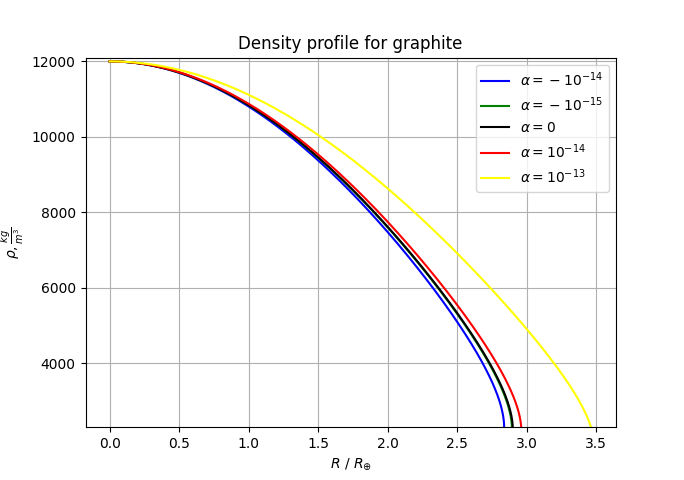}
\caption{[color online] Mass-radius relation and density profile plots for homogeneous cold spheres of graphite for different values of the parameter $\alpha$ ($\alpha=0$ gives a GR/Newtonian planet), obtained from the SKHM equation of state with the used of the modified hydrostatic equilibrium equations. }
\label{graphite}
\end{figure*}

Our results are given by the plots (\ref{fe})-(\ref{graphite}), that is, mass-radius relations and density profiles for the six most common solid materials found in rocky planets: iron, water ice, silicates, silicon carbide, and graphite. Black curves ($\alpha=0$) represent GR/Newtonian case.
The curves' flattening on the mass-radius plots (\ref{fe})-(\ref{graphite}), related to the constant planets' radii and even decreasing their values for larger masses, occurs because of electron degeneracy, whose pressure becomes important at high mass (see the discussion in \cite{zapol}).

We have also designated the positions on the figures for four rocky (the Earth, Venus, Mars and Mercury) and two ice-giant planets (Uranus and Neptune) of the Solar System, as well as a few exoplanets: super-Mercuries \cite{brugger1,nasa2} on the iron curves (\ref{fe}) and super-Earths \cite{nasa2,brugger2,ago} on the silicate curves (\ref{mgsio}) and (\ref{mgfe}). As we observe, the Earth-like planets can be found along the silicates curves while Mercury-like ones along the iron curves, because those materials are the most abundant in their interiors. Uranus and Neptune are situated much above the water-ice curves\footnote{They are found between water-ice and the helium curves, where the last one is not depicted here.} because of the high abundances of helium, which cannot be modelled by the used in this work equations of state \cite{zapol,seager}. Although our studies are related to toy-model planets, that is, we consider only homogeneous planets without taking into account multiple layers of different compositions, the most important result obtained via this analysis is an additional degeneracy in the mass-radius relations caused by modified gravity models. In the GR/Newtonian case, when one deals with a planet having multiple layers of varied compositions, different mass fractions of, for instance, iron cores and silicate mantles can provide the same total radius for the transiting planet of the same mass \cite{seager}. Apart from this, we have demonstrated that a similar degeneracy will appear when one applies modified structural equations. A curve of a given material obtained from different than the GR/Newtonian gravity model is shifted with respect to the GR/Newtonian curve and it can overlap with a mixture of two or more materials, whose curve was plotted using GR/Newtonian equations.

Moreover, let us notice that, even for the Birch-Murgnagham EoS considered in the above subsection, we are dealing with a notable difference between GR/Newtonian curves and modified gravity ones for the density profiles. That fact can be used to test and to constrain models of gravity when we are equipped with seismic data coming from earthquakes and marsquakes - see the discussion in the next section.

\section{A new test for models of gravity}\label{test}

As clearly demonstrated, although in the case of the simplified modelling of a planet as a single-material sphere, modified gravity affects the internal structure and density profile of such an object. This fact immediately equips us with the possibility to test models of gravity when the internal structure of the planet is well-known.

When the planet's density profile $\rho(r)$ is given, the polar moment of inertia $\mathcal{C}$ can be obtained by the expression:
\begin{equation}\label{polar}
    \mathcal{C}=\frac{8\pi}{3}\int_0^R \rho(r) r^4 dr, 
\end{equation}
where $R$ is the planet's radius. Roughly speaking, knowing a planet's profile means that the number of differently composed layers and their boundaries are provided. So far, the best known planet's inner structure is that of the Earth, endowed by the PREM model \cite{prem} and its further improvements \cite{kustowski,iasp91,aki135} (see more models at \cite{iris}), whereas we will be soon equipped with the Mars one given by the Seismic Experiment for Interior Structure (SEIS), that is,
from NASA's MARS
InSight Mission's seismometer \cite{nasa}.

The seismic wave changes when it travels through different layers of the planet. These changes depend on the material that the layer is made of, allowing to describe the material's characteristics, such as for example bulk modulus $K_0$ (incompressibility) which appears in the EoS (\ref{BME}) via the velocities of the longitudinal and transverse elastic waves, or via seismic parameter \cite{poi}.

These density profiles, given by the PREM model and its improvements, assume the Newtonian gravity. However, as revealed in our simplified case, the density profiles are also affected by the model of gravity, giving curves slightly different when compared to the ones obtained with the use of the Newtonian equations. We also suspect that the layers' thickness will be influenced by a gravity model applied when more realistic internal structure is used, that is, when one takes into account the different layers structure of a given planet. These facts will also have an influence on the polar moment of inertia (\ref{polar}), giving a different result for each model of gravity.

On the other hand, a normalized polar planet's moment of inertia $\mathcal{C}/MR^2$ ($M$ is a planet's mass) is a quantity which can be obtained from the relation for precession rate $d\psi/dt$ being caused by gravitational torques from the Sun \cite{kaula}
\begin{equation}\label{precession}
    \frac{d\psi}{dt}=-\frac{3}{2}J_2\cos{\epsilon}(1-e^2)\frac{n^2}{\omega}\frac{MR^2}{\mathcal{C}}
\end{equation}
where $e$ is the orbital eccentricity, $\epsilon$ the obliquity, $\omega$ the rotation rate, $n$ is the effective mean motion while $J_2MR^2$ is a factor consisting of the principal moments of inertia (see e.g. \cite{bill,folk}, $J_2$ is the gravitational harmonic coefficient). This factor, as well as the other mentioned quantities, is well-known in the case of the Solar System planets with high accuracy, in particular for the Earth \cite{ziemia} as well as for Mars  from the Viking \cite{konopliv,smith}, Mars Pathfinder \cite{folk2} and other missions. So far, the opposite procedure has been applied in order to get to know the internal structure of a planet, as, for instance, in the case of Mercury \cite{margot,brug,stein,harder,spohn,riner} and Venus \cite{margot2,konop,dumo}: from the many proposed density profiles and structural models survived only those which give the moment of inertia (\ref{polar}) compatible with the one provided by the accurate observational data (\ref{precession}).

In what follows, we propose a procedure which allows to constrain models of gravity using the already available seismic data of the Earth, and of Mars, when SEIS data obtained from analyzing waves created by marsquakes, thumps of meteorite impacts, surface vibrations generated by activity in Mars' atmosphere, and by weather phenomena, e.g. dust storms \cite{nasa}, are ready.
Such obtained density profiles in a given model of gravity, although sometimes carrying uncertainties regarding the most internal layers (the core), can be used to compute the polar moment of inertia (\ref{polar}) which must agree with the high accurate value acquired by observations (\ref{precession}). Although the figures (\ref{bm_fe}) and (\ref{bm_mgfe}) of the mass-radius relations  do not demonstrate deviations from Newtonian model in the case of the Earth and Mars, there are significant differences in the density profiles for the used values of the parameter $\alpha$. It will affect the polar moment of inertia (\ref{polar}).

\section{Discussion and conclusions}\label{concl}
In this work we have studied homogeneous cold spheres of low masses in the framework of Palatini $f(\mathcal{R})$ gravity. Those spheres are made of one of the six solid materials which are the most common substances found in the terrestrial planets. The mass-radius relations and density profiles obtained for each of the considered materials in our analysis have revealed that even for such low densities as the ones present in the rocky planets modified gravity changes the curves, allowing to draw interesting conclusions.

The mass-radius relation for homogeneous (and multiple layers' structure) spheres of various chemical compositions is an important tool providing an idea of the most abundant materials which an (exo-)planet consist of, when its mass and radius are known, mainly from the observations of transiting exoplanets. Although such observations can carry quite big uncertainties in the mass/radius ratios, when once discovered, more powerful telescopes follow up new exoplanets to get more precise data. Furthermore, we are living in a very exciting epoch - more and more current and future scientific missions, such as, for instance, current Cosmic Vision 2015-2025 (with a special focus on Cheops, Plato, Ariel, and Jupiter Icy Moons Explorer) \cite{vision} with further extensions of Voyage 2050 \cite{voyage} from ESA, or James Webb Space Telescope \cite{webb}, Nancy
Grace Roman Space Telescope \cite{nancy}, The Transiting Exoplanet Survey Satellite \cite{tess}, Spitzer Space Telescope \cite{spitzer}, and NN Explore \cite{nn} from NASA, are/will be collecting data on the Solar System planets and from other star's systems.

Apart from the findings in regards to (exo-)planets discussed in more detail below, we have also examined carefully a possible singular behaviour of our equations, caused by the extra terms derived from Palatini quadratic model (especially the one related to the conformal transformation \cite{aneta,olek}). More precisely, an eventual ill behaviour of the hydrostatic equilibrium equations (\ref{tovJ}), leading to a non-physical behaviour of a spherical-symmetric system such as a planet or a star, could appear for a certain value of the parameter $\alpha$. This particular value, as we observe from the relation (\ref{singu}), is related to an equation of state. This is the reason why one needs to be careful when choosing the range of the parameter $\alpha$ (in the negative values' part) such that the considered EoS will not produce those values.

Concluding, our results can be summarized to the following three main points:
\begin{itemize}
    \item {\bf Extra degeneracy in the mass-radius relation}\newline
    Apart from the well-known degeneracy in the mass-radius relation, making the determination of the numbers of layers and their properties problematic already in GR/Newtonian model \cite{seager}, modified gravity introduces another one, caused by the additional theory parameter. Therefore, a transiting exoplanet may have slightly different layer structure and composition of each of the layers than the ones predicted by the GR/Newtonian model, especially in the case of carbon, water and silicate planets (see the next point). Moreover, since iron is the most dense element out of which a planet can form, exoplanets with radii smaller than pure iron planets are not expected to be found. This limit is also well-known, but as we can see in the figure (\ref{fe}), modified gravity may shift the curves in the smaller radii region. Finding more exoplanets of the Mercury's type with very small radius \cite{submerkury} could be an additional indication that other model than General Relativity and its Newtonian limit may have something to say in the planetary physics.
    \item {\bf Exoplanets properties}\newline
    There are a few interesting properties one may tell about a type of transiting exoplanets when their masses and radii are known - the smaller uncertainties in the mass and radius estimations\footnote{which are related to the host star' properties which also depend on theory of gravity; for more details, see \cite{early,Wojnar:2020frr}.}, the more characteristics of the exoplanet can be given. For example, as demonstrated in \cite{seager}, planets above the water ice curve must be richly abundant in the hydrogen/helium in their envelope, therefore super-Earth exoplanets do not possess significant gas envelopes. As shown in our figure (\ref{h2o}), modified gravity may alter a bit this conclusion, since the large planets which are believed to have a considerable amount of those light elements in their atmospheres could be very poorly equipped in them in the case of other than GR/Newtonian theory of gravity.
    
    Moreover, modified gravity can also shed light on some of the already existing hypothesis/problems related to the planetary physics, as for example a planet migration and our knowledge on a planetary system formation. Planets of a certain structure and compositions are expected to be found at a particular distance from their host stars. Water planets (= planets with more than $25\%$ water ice by mass) can be identified with up to $5\%$ fractional uncertainty in mass and radius and tend to have large radii \cite{valencia}. They are usually found far from their host stars and that fact, together with the current model of planetary system formation, is one of the main leading point standing behind the idea of planet migrations if the water planets would be found on a closer orbit. Therefore, one rather expects to detect a water exoplanet at an edge of a certain host star's system, while the Earth-like, with the silicate main contribution, are supposed to be found in much smaller distance. This proposition is derived from the special position of the water ice curve on the mass-radius plots, being an important hint for distinguishing rocky exoplanets from water ones and those with a rich in helium and hydrogen atmospheres.  However, as already marked out in the previous point, the water ice curve is shifted in modified gravity. The (exo-)planets lying close to the GR/Newtonian water curve may an reality be still solid planets, much less abundant in water and not possessing envelopes rich in the light elements. 
  
    \item {\bf Constraining models using seismic data from the Earth and Mars}\newline
    The significant difference in the density profiles (\ref{fe})-(\ref{graphite}) between GR/Newtonian model and ones provided by modified gravity, even for low-mass planets (\ref{bm_fe})-(\ref{bm_mgfe}), gives us an excellent opportunity to test and to constrain the existing models of gravity with the use of the known Earth' and near-future Mars' internal structures. The physics of low pressures and temperatures (in the meaning of planetary regimes) is much better understood than physics of stars and their compact and ultracompact remnants\footnote{although it is not free from issues such as, e.g., convective processes, inner core description, discontinuities between layers, to mention just few of them \cite{earth,earth2}.}. In the section \ref{test} we have described the procedure which, after improving our toy-planet's models, will be used to constrain given models of gravity. The polar moments of inertia for the Solar System planets, in particular, the Earth and Mars, are known with high accuracy (\ref{precession}), allowing to compare with the ones obtained from seismic data and the model of gravity, and eventually to discard those being inconsistent with the observed one.
    
    Let us notice that our studies predict significant differences in the layer's thickness even for small values of the parameter $\alpha$. They are more noticeable in the case of the heavier elements. Consequently, we speculate that the most inner layers, such as cores and mantles, consisting of mainly iron and silicates, respectively, will vary in modified gravity models in comparison to the current ones. However, in order to say more in regards to that topic, we need to improve our model enriching it in multiple layers of different compositions as suggested by the PREM and other models, as a starting point. Subsequently, more consistent approach will require a re-examination of the current Earth and Mars models using earthquakes' and marsquakes' data, together with modified structure equations. Therefore, when seismic data are used in order to obtain density profiles in a given model of gravity - that is, when we know well enough the materials composing the different layers of the planet as well as the layers' thickness in GR/Newtonian and the gravity model, we may use it to examine the deviations caused by various gravitational proposals. Such a procedure will enable us to constrain the modified gravity models with respect to the given accuracy.
    
    Let us just mention, as a concluding remark to that point, that so far only helioseismic data has been used to constrain modified gravity models \cite{casa, saltas}.
\end{itemize}

The results presented here and followed by the future investigations will provide an accurate test and constraints of models of gravity from our nearest playground: the Earth and Mars. Works regarding this topic are already under development and will be soon presented to the physics community.

\vspace{5mm}
\noindent \textbf{Acknowledgement.} 
This work was supported by the EU through the European Regional Development Fund CoE program TK133 ``The Dark Side of the Universe." AK is a beneficiary of the Dora Plus Program, organized by the University of Tartu.\\
The authors would like to thank Mar\'ia Jos\'e Vera and Gerardo Tejada Saracho for enlightening discussions on seismology and Mars' missions.

\appendix
\section{Structural equations used in the numerical analysis}\label{appen}

For the SKHM (modifed polytropic) equation of state
\begin{equation}
      \rho(p)=\rho_0 +cp^n,
\end{equation}
the Palatini hydrostatic equilibrium equation (\ref{tovJ1}) used in the numerical analysis has the following form
{\scriptsize
\begin{equation}\label{rel}
    \begin{split}
       & \frac{d \rho}{d r}=   -\frac{ G \mathcal{M} \rho  \left(\frac{4\pi
   \alpha  \rho ^2 r^3}{\mathcal{M} (2\alpha  \rho +1)^{1/2}}+1\right)}{r^2 (2 \alpha  \rho
   +1)^{1/2} \left(1\, -\frac{2 G \mathcal{M}}{c^2 r (2 \alpha  \rho +1)^{1/2}}\right)
   } \\
   & \times \frac{1}{\left(\frac{ \alpha  G \mathcal{M} \rho  \left(\frac{4\pi \alpha  \rho ^2 r^3}{\mathcal{M}
   (2 \alpha  \rho +1)^{1/2}}+1\right)}{r (2 \alpha  \rho +1)^{3/2} \left(1\,
   -\frac{2 G \mathcal{M}}{c^2 r (2 \alpha  \rho +1)^{1/2}}\right)}+\frac{2\alpha  c^2 \rho
   }{2\alpha  \rho +1}+\frac{\left(\frac{\rho - \rho_0}{c_p}\right)^{\frac{1}{n}-1}}{c_p n}\right)}
    \end{split}
\end{equation}
}
while the mass function is given by the equation (\ref{masaRel}).

Let us notice that when we consider a non-relativistic limit of the full relativistic equation (\ref{tovJ}), one performs the expansion around $\alpha=0$. This procedure causes the sign change in the extra term related to the modification in the non-relativistic equation, which for the given above EoS takes the form:
\begin{equation}\label{wrong}
    \frac{d\rho}{dr}\approx -\frac{Gm(r)\rho}{r^2}\frac{ cn\left(\frac{\rho-\rho_0}{c}\right)^\frac{n-1}{n}(1-\alpha\rho)}{\left(1+2\alpha \frac{Gm}{r}cn\rho\right)\left(\frac{\rho-\rho_0}{c}\right)^\frac{n-1}{n}}.
\end{equation}

This is misleading and leads to the non-physical behaviour of the mass-radius and density curves in the case of large masses (large densities): in such a situation the term $(1-\alpha\rho)$ may take negative values, changing the sign of the full expression (\ref{wrong}). It also suggests that a singular behavior of the equations would happen for positive values of the parameter $\alpha$, but as we have already discussed in the subsection \ref{alfa}, an eventual singularity occurs for negative ones. However, although we have not demonstrated it here, the non-relativistic equation (\ref{wrong}) can be used for low-masses planets, with $M/M_\text{Earth}<5$. In order to consider (exo-)planets' higher masses as well, we have been using the equation (\ref{rel}), where we just skipped pressure terms in the combinations with energy density $\rho+p/c^2\approx\rho$ and mass $pr^3/c^2\mathcal{M}\approx0$ as its influence is insignificant for the objects we study here.

\end{document}